\documentstyle[12pt]{article}

\newcommand{\be}{\begin{eqnarray}}
\newcommand{\ee}{\end{eqnarray}} 

\begin{document} 

\begin{titlepage}

\hrule 
\leftline{}
\leftline{Preprint
          \hfill   \hbox{\bf CHIBA-EP-97}}
\leftline{\hfill   \hbox{hep-ph/9708260}}
\leftline{\hfill   \hbox{May 1997}}
\vskip 5pt
%\leftline{}
\hrule 
\vskip 1.0cm

\centerline{\large\bf 
Running Flavor Number and Asymptotic Freedom
} 
\centerline{\large\bf  
{
in the Normal Phase of QED
}$^*$
}
   
\vskip 0.5cm

\centerline{{\bf 
Kei-Ichi Kondo$^{1,2}{}^{\dagger}$
and Takeharu Murakami$^2{}^{\ddagger}$
}}  
\vskip 4mm
\begin{description}
\item[]{\it  %$^1$ 
$^1$ Department of Physics, Faculty of Science,
  Chiba University, Chiba 263, Japan
  %$^\ddagger$
  }
\item[]{\it 
$^2$ Graduate School of Science and Technology,
  Chiba University, Chiba 263, Japan
  %$^\ddagger$
  }
\item[]{$^\dagger$ 
  E-mail:  kondo@cuphd.nd.chiba-u.ac.jp 
  }
\item[]{$^\ddagger$ 
  E-mail:  tom@cuphd.nd.chiba-u.ac.jp  
  }
\end{description}
\vskip 0.5cm

\centerline{{\bf Abstract}} \vskip .5cm

In the normal phase (where no dynamical fermion mass
generation occurs) of the $D$-dimensional  quantum
electrodynamics with $N_f$ flavors of fermions,
we derive an integral equation which should be satisfied by
(the inverse of) the wave function renormalization of the
fermion in the Landau gauge. For this we use the inverse
Landau-Khalatnikov transformation connecting the
nonlocal gauge with the Landau gauge. This leads to
a similar equation for the running flavor
number in the framework of the $1/N_f$ resumed
Schwinger-Dyson equation. Solving the equation
analytically and numerically, we study the infrared behavior
and the critical exponent of the 3-dimensional QED
(QED$_3$). This confirms that the flavor number in
QED$_3$ runs according to the
$\beta$ function which is consistent with the asymptotic
freedom as that in 4-dimensional QCD.
\vskip 0.5cm
%Key words:

%PACS:  
\vskip 0.5cm
\hrule  
%%%%%%%%%%%%%%%%%%%%%%%%%%%%%%
%\vskip 2cm  
%\hrule  
%\bigskip  
%\centerline
%{\bf CHIBA UNIVERSITY}  
%\vfill 

\vskip 0.5cm  
%\hrule  

%\begin{description}
%\item[]{
%$^\ddagger$
%Address from March 1996 to December 1996.
%  On leave of absence from: \\
%  Department of Physics, Faculty of Science,
%  Chiba University, Chiba 263, Japan.
%  }
%\item[]{
$^*$ To be published in Physics Letters B.
% Submitted to .
% }  
%\end{description}

\end{titlepage}

%\newpage
%%%%% Table of Contents %%%%%
%\pagenumbering{roman}
%\tableofcontents
%%%%% Table of Contents %%%%%
\pagenumbering{arabic}
%%%%%

\newpage
%\section{Introduction}
%\setcounter{equation}{0}
\par
Several years ago \cite{KN92} running of the flavor
number $N_f$ has been investigated in the three-dimensional
QED (QED$_3$) based on the $1/N_f$ improved Schwinger-Dyson
(SD) equation. It was shown that the parameter $1/N_f$
exhibits the asymptotic freedom in QED$_3$ as in
non-Abelian gauge theory in 3+1 dimensions.  This
conclusion was drawn by solving the SD equation for
the wave function renormalization for the fermion in the
Landau gauge including the vertex correction.  In order to
arrive at the above conclusion, however, we needed to take
several approximations.  They are reexamined 
and some of them are criticised in \cite{AAKMM96}.  In the
previous work 
\cite{Kondo97} the running of the flavor number was
translated into the behaviour of the nonlocal gauge which
can be derivable for a more general class of the vertex
function than that considered in \cite{KN92}.  This work
has resolved the criticism raised by \cite{AAKMM96}.  
\par
The purpose of this Letter is to rederive the running of
the flavor number based on a different method which is
superior to the previous one \cite{KN92} in the sense
that the new method is free from the various ambiguities and
most of the approximations taken in the previous paper
\cite{KN92}.  
This result enables us to understand the normal phase of
high-T$_c$ superconductor \cite{BM86} as a non-trivial IR
fixed point, see
\cite{AM96}. 
The equation derived in this Letter provides a useful tool
to study, among other things, the existence of non-trivial
quasi infrared fixed point and slowing down of the running
flavor number in the intermediate region, which is
responsible for the deviation from the Fermi-liquid
behavior in the normal phase of high-$T_c$
superconductivity.

\par
We focus on the SD equation for the full fermion
propagator, 
$
S(p) = [A(p^2)\hat{p}-B(p^2)]^{-1},
$
in D-dimensional QED.
The SD equation for $S(p)$ is decomposed into a pair of
two coupled nonlinear integral equations for $A(p^2)$ and
$B(p^2)$.  In general, this procedure leads to rather
complicated coupled equations.
A method of avoiding the
coupled equation is to find a situation such that the
wave function renormalization disappears,
i.e. $A(p^2) \equiv 1$ (no wave function renormalization).
This is realized in the framework of SD
equation. 
\footnote{
In perturbation theory this is possible order by
order in the coupling constant. 
}
Actually, under the bare vertex approximation, the fermion
wave function renormalization becomes 1 identically in the
Landau gauge in any dimension $D>2$, if the bare photon
propagator is adopted in the SD equation of QED
(i.e. in the quenched approximation), see e.g. 
Ref.~\cite{KN89}. This extremely simplifies the
actual analysis of SD equation, since we have only to solve
a single equation for $B$
(See Ref.~\cite{Kondo97} for more advantages of the
nonlocal gauge). This scenario has been extended beyond the
quenched approximation
\cite{GSC90,KM92,KEIT95,KM95,Kondo97}.  Especially, it has
been shown
\cite{Kondo97}  that  in $D$-dimensional QED
we can choose a covariant {\it nonlocal} gauge in which 
$A(p^2) \equiv 1$ follows as a solution of the coupled SD
equations, if the vertex function belongs to the following
class:
\begin{eqnarray}
 \Gamma_\mu(p,q; k) = \gamma_\mu G(p^2, q^2, k^2) ,
\end{eqnarray}
where $G$ is an arbitrary function of the arguments except
that $G$ does not depend {\it explicitly} on
$k^2$ (Note that implicit dependence on $k^2$ through $A(k)$
is allowed \cite{Kondo97}).
Such a nonlocal gauge 
\footnote{
This is obtained by integrating the first order
differential equation satisfied by $\xi$.  Therefore, we
need a boundary condition, see
Ref.~\cite{GSC90,KM92,Kondo97}.
}
$\xi =\xi(k)$
(i.e. momentum-dependent covariant gauge fixing, see
Ref.~\cite{KM95,Kondo97} for the Lagrangian formulation)
is given by 
\begin{eqnarray}
\tilde \xi(k^2) := \xi(k)[1-\Pi(k)/k^2]
 &=& 1 + {D-2 \over (k^2)^{D-1} D_T(k^2)}
\int_0^{k^2} dz \ z^{D-1} {d D_T(z) \over dz} ,
\label{nlg}
\end{eqnarray} 
when the full photon (or gauge boson) propagator has the
form
\begin{eqnarray}
 D_{\mu\nu}(k) 
 &=& D_T(k^2) \left( g_{\mu\nu} - {k_\mu k_\nu \over k^2}
\right)
 + {\xi(k) \over k^2} {k_\mu k_\nu \over k^2},
 \nonumber\\
 D_T(k^2) &:=& {1 \over k^2 - \Pi(k)} ,
 \label{gbpropa}
\end{eqnarray}
where $\Pi(k)$ is the photon vacuum polarization function.
In fact, in the quenched  
($\Pi(k) \equiv 0$) and ladder ($G \equiv 1$) approximation,
eq.~(\ref{nlg}) recovers the Landau gauge
$\xi(k) \equiv 0$ for $D>2$.
Therefore, eq.~(\ref{nlg}) is indeed a generalization of the
quenched case. As a result,
we have only to solve a single equation for
$B$:
\begin{eqnarray}
 B(p^2) &=& m_0 + e^2 \int {d^Dq \over (2\pi)^D} {B(q^2)
G(p^2, q^2, k^2) \over q^2 +B^2(q^2)}  D_T(k^2)
[D-1+ \tilde \xi(k^2)] .
\label{SDBnlg}
\end{eqnarray} 
The mass function $M(p):=B(p^2)/A(p^2)$ reduces simply to
$B(p^2)$ in the nonlocal gauge (\ref{nlg}).
The SD equation (\ref{SDBnlg}) (without
IR cutoff) has been solved for QED$_3$ analytically and
numerically in
\cite{KEIT95,KM95}.  
\par
A question is how one can recover the Landau
gauge result in this scheme.  As remarked in the previous
work
\cite{Kondo97}, the {\it inverse} Landau-Khalatnikov (LK)
transformation
\cite{LK56} can recover the full fermion propagator
$S(x)$, the full gauge boson propagator $D_{\mu\nu}(x)$
and the full vertex function ${\cal V}_\mu(x,y,z)$ in the
Landau gauge from those in the nonlocal gauge (\ref{nlg}).
\par
The LK and the inverse LK transformations leave the system
of the SD equation {\it form invariant}.
Usually the LK transformation is used to transform the
results obtained in the Landau gauge into those in other
gauges.  This is because, in many cases, the calculation in
the Landau gauge is easier than other gauges. (This
is not necessarily true in the study of SD equation.) 
However, it is easy to see that the LK transformation is
invertible, see Ref.~\cite{Kondo96b}.  Therefore the
inverse LK transformation enables us to transform the
results in the arbitrary gauge (even the nonlocal gauge)
into those in the Landau gauge. 
\par
For example, in the $D=d+1$ dimensional QED the fermion
propagator
$S_{NLG}(x)$ in the nonlocal gauge is connected with that
$S_L(p)$ in the Landau gauge (in configuration
space)
\cite{Kondo97}.
\begin{eqnarray}
  S_L(x) = e^{-\Delta(x)} S_{NLG}(x) ,
 \label{LKnlg}
\end{eqnarray}
where
\begin{eqnarray}
 \Delta(x) :=  e^2 \int {d^Dk \over (2\pi)^D} 
 (e^{ik \cdot x}-1) f(k),
 \label{delta}
\quad f(k) := {\xi(k) \over k^4} .
 \label{fk0}
\end{eqnarray}
Especially, in the normal phase (with no fermion mass
generation, i.e. $B(p^2) \equiv 0$), $S_{NLG}(x)$ agrees
with the bare fermion propagator $S_0(x)$:
\begin{eqnarray}
 S_0(x) 
 = \int {d^Dp \over (2\pi)^D} e^{i p \cdot x}
 {\gamma^\mu p_\mu \over p^2}
 = i \gamma^\mu x_\mu {\cal P}_0(x),
 \quad 
 {\cal P}_0(x) = {\Gamma(D/2) \over 2\pi^{D/2}|x|^{D}}.
 \label{freeSd}
\end{eqnarray}
Therefore, the momentum-space propagator $S_L(p)$ in the
Landau gauge
\begin{eqnarray}
 S_L(p) = {1 \over A(p^2)\gamma^\mu p_\mu} 
 \label{fpL}
\end{eqnarray}
is
obtained by applying the Fourier transform to
(\ref{LKnlg}).  Thus the wave function renormalization in
the Landau gauge is obtained in the normal phase 
\cite{Kondo97},
\begin{eqnarray}
Z(p) := A^{-1}(p^2) 
= i \int d^Dx e^{i p \cdot x} e^{-\Delta(x)} 
 (p \cdot x) {\cal P}_0(x) .
 \label{wfrL}
\end{eqnarray}
However, in order to obtain the wave function
renormalization in the Landau gauge according to
(\ref{wfrL}), we need to perform the Fourier integration
twice (in (\ref{delta}) and (\ref{wfrL})).  Although this
is not impossible in principle, it becomes rather laborious
to perform it when the nonlocal gauge
$\xi(k)$ is a complicated function of $k$
(This is indeed the case of QED$_3$ in question).

\par
In this paper we propose another method which is superior
to the above method (\ref{wfrL}) as shown in what follows.
A basic observation is that in the normal phase the function
$Z(p):=1/A(p^2)$ in the Landau gauge satisfies the
following integral equation,
\begin{eqnarray}
  Z(p) = 1 + i e^2 \int {d^D q \over (2\pi)^D}
  {q \cdot (p-q) \over q^2} f(p-q) Z(q) ,
  \label{eqZ0}
\end{eqnarray}
where $f(k)$ is given by (\ref{delta}). 
This equation is derived as follows.
Note that in the normal phase $S_{NLG}(x)$ is equal to
the massless bare fermion propagator $S_{0}(x)$ which
satisfies
$i \hat{\partial}S_{0}(x)=\delta^D(x)$
where $\hat{\partial}:=\gamma^\mu \partial_\mu$.
Operating the differential operator $\hat{\partial}$ to
both sides of (\ref{LKnlg}), therefore, we obtain
\begin{eqnarray}
  i \hat{\partial} S_L(x) = \delta^D(x) 
  - ie^2 S_L(x) \int {d^Dk \over (2\pi)^D}  \hat{k} 
  f(k)  e^{-i k \cdot x} ,
  \label{SDc}
\end{eqnarray}
where we have used a fact $\Delta(0)=0$.
Moving from the
configuration space to the momentum space, we obtain
\begin{eqnarray}
  \hat{p} S_L(p) = 1 + ie^2 \int {d^Dk \over (2\pi)^D}
  S_L(p-k) \hat{k}  f(k) ,
\label{SDt}
\end{eqnarray}
since the Fourier transformation changes the simple
product in coordinate space into the convolution in
momentum space.
By substituting (\ref{fpL}) 
into (\ref{SDt}), we arrive at the result 
eq.~(\ref{eqZ0}) after the change of
variable $q=p-k$.
\par
We can interpret the integral
equation (\ref{eqZ0}) as a SD equation derived from the
fermion propagator $S$ itself, not $S^{-1}$.
The conventional SD equation derived from
$S^{-1}(p)=A(p^2)\hat{p}-B(p^2)$ in momentum space leads to
the SD equations for $A$ and $B$, which are nonlinear
integral equation in $A$ and $B$.
The exact solubility of eq.(\ref{eqZ0}) is due to the
linearity of the equation in $Z(p)$.
It is worth remarking that the SD equation (\ref{eqZ0}) can
be derived from another type of the vertex function
\cite{Kondo96b}. See Ref.~\cite{Kondo96b} for more details.

\par
In order to solve the equation (\ref{eqZ0}), then, we move
to the Euclidean space and separate the integration
variables.  Then we arrive at the equation,
\begin{eqnarray}
  Z(p) = 1 -  e^2 \int_{\epsilon}^{\Lambda} dq
Z(q) L_D(p,q) , 
\label{eqZ}
\end{eqnarray}
where $p:=\sqrt{p^2}$, $q:=\sqrt{q^2}$ and the kernel for
$D>2$ is given by
\begin{eqnarray}
  L_D(p,q) := {q^{D-3} \over 2^{D-1}
\pi^{(D+1)/2}\Gamma({D-1 \over 2})}
  \int_0^\pi d \vartheta \sin^{D-2}\vartheta 
  (pq \cos \vartheta - q^2) f(\sqrt{(p-q)^2}) .
  \label{kernel}
\end{eqnarray}
Here we have introduced the ultraviolet (UV) cutoff
$\Lambda$ and the infrared (IR) cutoff $\epsilon$.
\par
In the following we restrict our arguments to the
 multiflavor QED$_3$.
In QED$_3$ the {\it dimensionful} quantity $\alpha$ can be
used as an effective UV cutoff of the momentum integral,
which is supported by the Wilsonian approach of the
renormalization group, see \cite{Kondo97} and references
therein.  Hence we put
$\Lambda=\alpha$ hereafter. (Application of eq.(\ref{eqZ0})
to other models will be presented elsewhere
\cite{KM97}.)
In the normal phase, the vacuum polarization (at one
loop) reads
\begin{eqnarray}
 \Pi(k) = - \alpha k, \quad \alpha := N_f e^2/8.
\end{eqnarray}
Note that this quantity is gauge invariant.
\par
The nonlocal gauge in the absence of IR cutoff 
\cite{KEIT95,Kondo97} is given by
\begin{eqnarray}
  \tilde \xi(k^2) = 2 - 2 {k^2+\alpha k \over k^4}
  \left[ k^2 - 2 \alpha k + 2\alpha^2 \ln \left( 1+{k
\over \alpha} \right) \right] .
\label{nlgexpa}
\end{eqnarray}
Then we obtain
\begin{eqnarray}
  f(k) = {2 \over k^4 + \alpha k^3} - {2 \over k^4} \left[
  1 - {2\alpha \over k} + {2\alpha^2 \over k^2} 
  \ln \left( 1+{k \over \alpha} \right) \right] .
  \label{fk}
\end{eqnarray}
This should be substituted into the kernel given by
\begin{eqnarray}
  L_3(p,q) := {1 \over 4 \pi^2}
  \int_0^\pi d \vartheta \sin \vartheta 
  (pq \cos \vartheta - q^2) f(\sqrt{(p-q)^2})  .
  \label{3dK}
\end{eqnarray}
Note that the kernel $L_3(p,q)$ is not symmetric in 
$p, q$, i.e. $L_3(p,q) \not= L_3(q,p)$. The angular
integration in the kernel
$L_3(p,q)$  can be done exactly, see Appendix.  However,
the expression is rather involved. The full analysis of
equation (\ref{eqZ}) with an exact kernel $L_3(p,q)$ will
be done using the numerical method later. 
\par
First, we give an analytical solution by taking into
account the leading term only in the IR expansion:
\begin{eqnarray}
  f(k) 
  = \sum_{n=0}^{\infty} 2 (-1)^n {n+1 \over n+3} 
  {k^{n-3} \over \alpha^{n+1}} 
  =  {2 \over 3\alpha k^3} - {1 \over \alpha^2 k^2}
  + O({1 \over k}) ,
  \label{expandf}
\end{eqnarray}
and then discuss the effect of the other terms.
It should be remarked that the factor $2/3$ in the leading
term of $f(k)$ comes from the IR value 
$\tilde \xi(0)=2/3$ of the nonlocal gauge (\ref{nlgexpa}) in
the absence of IR cutoff \cite{KEIT95}:
\begin{eqnarray}
  \tilde \xi(k^2) = {2 \over 3} - {k \over 3\alpha} 
  +  O(k^2) .
\end{eqnarray}
Up to the leading term in (\ref{expandf}), the kernel
reduces to a very simple form:
\begin{eqnarray}
   L_3^{(1)}(p,q) =  - {1 \over \pi^2}
    {1 \over 3\alpha q}\theta(q-p) ,
\label{3dK1}
\end{eqnarray}
where $\theta(x)$ is the Heaviside step function.
%\footnote{
%This should be compared with the case
%$
%  f(k) = {\xi \over k^4} .
%$
%In this case, the kernel reads
%$
%  K(x,y) =  .
%$}
The integral equation (\ref{eqZ}) with the kernel
(\ref{3dK1}) can be converted to the boundary value
problem of the first order differential equation:
\begin{eqnarray}
 {1 \over Z(p)} p {d \over d p} Z(p) = - \gamma  ,
\quad
 \gamma := {8 \over 3\pi^2 N_f} .
 \label{ad}
\end{eqnarray}
The general solution of the differential equation is given
by
$
 Z(p) = C p^{-\gamma}
$
with a constant $C$.
Imposing the UV boundary condition $Z(\alpha)=1$ at
$p=\alpha$, the solution is uniquely determined:
\begin{eqnarray}
  Z(p) 
  = \left({p \over \alpha}\right)^{-\gamma} ,
  \quad
  \gamma := {8 \over 3\pi^2 N_f} .
  \label{wfr}
\end{eqnarray}
Note that $Z(p) \uparrow +\infty$ 
(or $A(p^2) \downarrow 0$) as $p \downarrow 0$.
The solution (\ref{wfr}) can also be obtained directly
using the iterated kernel as follows.
\footnote{
Using the kernel
$
K(p,q) = - e^2 L_3^{(1)}(p,q) = (\gamma/q) \theta(q-p) ,
$
Eq.~(\ref{eqZ}) is replaced by
$
  Z(p) = 1 + \int_{0}^{\Lambda} dq
 K(p,q) Z(q) . 
$
Then the solution is given by
$
Z(p) = 1 + \sum_{n=1}^{\infty} (K_n \cdot 1)(p),
$
where  
$
(K_n \cdot 1)(p) = \int_{0}^{\Lambda} dq
K(p,q) (K_{n-1} \cdot 1)(q) 
$
for $(n \ge 2)$,
and 
$
(K_1 \cdot 1)(p) = \int_{0}^{\Lambda} dq K(p,q) {\bf 1}(q)
=  \gamma \ln (\alpha/p) .
$
In this case, 
$
(K_n \cdot 1)(p)
= {\gamma^n \over n!} ( \ln {\Lambda \over p} )^n .
$
}
\begin{eqnarray}
  Z(p) =  1 + \sum_{n=1}^{\infty} {1 \over n!} 
  (\gamma \ln {\alpha \over p} )^n
  = \exp \left[\gamma \ln {\alpha \over p} \right] ,
  \label{wfr2}
\end{eqnarray}
which agrees with (\ref{wfr}).
The wave function renormalization in the IR regime
(\ref{wfr}) has been conjectured   based on either
theoretical speculations in
\cite{AH81,PW88,AJM90} or in numerical treatment
\cite{Maris96}.
Quite recently,  essentially the same result as (\ref{wfr})
was analytically obtained independently in \cite{AMM96} by
calculating (\ref{wfrL}) following the suggestion 
\cite{Kondo97}.
\footnote{
Note that the prefactor in $Z(p)$ is exactly 1
irrespective of
$N_f$ in (\ref{wfr}), while the
complicated dependence on $N_f$ appeared in \cite{AMM96}.
}
\par
The validity of the IR expansion used above has been
confirmed in the superconducting phase where dynamical mass
generation occurs \cite{KEIT95,KM95}.
However, the validity in the normal phase has not been  
confirmed yet.
Note that the solution (\ref{wfrL}) is independent of the IR
cutoff
$\epsilon$ due to the special nature of the kernel
(\ref{3dK1}). This property is not preserved if we include
the next-to-leading term in (\ref{expandf}), since the
kernel is modified as
\begin{eqnarray}
  L_3^{(2)}(p,q) =  {1 \over 4\pi^2} \left[
    {-4 \over 3\alpha q}\theta(q-p)
    + {1 \over \alpha^2} - {1 \over \alpha^2}
    {p^2-q^2 \over 2pq} \ln {p+q \over |p-q|} \right].
    \label{3dK2}
\end{eqnarray}
Therefore, the analysis at least up to the
next-to-leading order is necessary to see a possible
effect of the IR cutoff, in light of the discussion on the
cutoff dependence given in
\cite{KN92} where the ratio $\alpha/\epsilon$ plays an
essential role in the critical behavior. 
\par
An advantage of the method proposed in this paper is that
we can take into account all the terms up to any order in
a very similar way to the conventional SD equation.
This will be indispensable for studying the intermediate
region:
$\epsilon \ll p \ll \alpha$, which is responsible for the
deviation from the Landau Fermi-liquid behavior and the
existence of the non-trivial quasi IR fixed point
\cite{AM96}.
\par
A further advantage of this method is that we can
directly write down the equation for the running flavor
number (as a coupling constant) which was introduced in
\cite{KN92}

\begin{eqnarray}
  g_R(p) := g_0/A(p^2) = g_0 Z(p),
  \quad
  g_0 := {8/\pi^2  \over N_f} ,
\end{eqnarray}
where $g_0$ is the bare coupling constant.
Eq.~(\ref{eqZ}) implies that the running flavor number
in QED$_3$ obeys
\begin{eqnarray}
  g_R(p) = g_0 -  {\alpha \over 4} g_0
\int_{0}^{\alpha} dq L_3(p,q) g_R(q) .
\label{eqc}
\end{eqnarray}
Up to the leading term, the running coupling is obtained as
\begin{eqnarray}
  g_R(p)
  = g_0 \left({p \over \alpha}\right)^{-g_0/3}
  = g_0 \exp \left( -{g_0 \over 3} \ln {p \over \alpha}
\right).
  \label{rc1}
\end{eqnarray}
The study of the equation (\ref{eqc}) in the intermediate
region will shed light on the existence of the non-trivial
quasi IR fixed point of QED3 through  the $\beta$-function
\begin{eqnarray}
 \beta(g_R) =  {d \over dt} g_R(t), 
 \quad
 t := \ln (p/\alpha) ,
\end{eqnarray}
as well as the slowing down of the running in the
intermediate region.
\par
In contrast to the optimistic observation of
\cite{AMM96},  it is by no means obvious
whether or not the leading results survive the inclusion of
the higher order terms.
Therefore, we have solved Eq.~(\ref{eqZ}) for
$D=3$ numerically for the three kernels: the leading
(\ref{3dK1}), the next-to-leading (\ref{3dK2}) and the full
kernel (\ref{kernelL}) as given in the
Appendix. The numerical result is given in Fig.1 when
$N_f=5$ which lies in the normal phase (see the final
comment). Fig.1 shows that the solution (\ref{wfr})
in the leading order is a rather good approximation of the
full solution of Eq.~(\ref{eqZ}) in the IR region 
$p/\alpha < 0.1$. In the IR region the full solution is
approximated by (\ref{wfr}) with the same $\gamma$.
The IR critical exponent $\gamma$ defined by (\ref{ad}) is
nothing but the anomalous dimension. 
Therefore it turns out that the result of \cite{AMM96} is in
fact correct with regard to the critical exponent $\gamma$,
the result for
$A$ being changed merely by a multiplicative constant.
\par
In conclusion our method confirms that the leading solution
(\ref{rc1}) for the running flavor is also a good
approximation of the full solution of (\ref{eqc}).   We
find that Eq.(\ref{rc1})  is rewritten as
\begin{eqnarray}
  g_R(p)
  = g_0 [ 1 -{g_0 \over 3}\ln {p \over \alpha} + O(g_0^2)]
  = g_0/(1+ {g_0 \over 3}\ln {p \over \alpha}) + O(g_0^2),
  \label{app}
\end{eqnarray}
which implies the $\beta$-function:
\begin{eqnarray}
 \beta(g_R) =  - {1 \over 3} g_R^2 + O(g_R^3) .
\end{eqnarray}
Thus QED$_3$ exhibits the asymptotic freedom for the
coupling
$g_0$, at least for small $g_0$, as claimed in \cite{KN92}.
\par
In the nonlocal gauge, the running coupling is related to
the nonlocal gauge as \cite{Kondo97}
\begin{eqnarray}
  g_R^{NLG}(t) = g_0 \left[ 1 + {1 \over 2} \tilde \xi(k^2)
\right], \quad t := \ln (k/\alpha) .
\end{eqnarray}
This means that the growth of the running coupling is
cut off at the IR limit $k \rightarrow
0$, since $\tilde \xi(0)$ is finite irrespective of the
existence or absence of IR cutoff, in sharp contrast to
the result (\ref{rc1}), i.e.
$g_R(p) \uparrow +\infty$ as $p \downarrow 0$.  
Therefore, it is interesting to
see the effect of the covariant IR cutoff
\cite{Kondo97} in the Landau gauge and to see whether the
situation  change essentially or not.
In a subsequent paper \cite{KM97} we will give more 
detailed analysis on the running coupling and the fixed
point in the presence of covariant IR cutoff.  For this we
need to extend the LK transformation so as to include the
massive gauge field.

\par
Finally, we comment on the superconducting 
(chiral-symmetry breaking) phase where the dynamical fermion
mass generation occurs. It is worth
remarking that Appelquist et al.
\cite{ABKW86} assumed 
$A(p^2) \equiv 1$ in the Landau gauge from the
beginning based on the naive $1/N_f$ argument.
This became an origin of the criticism later by
Pennington et al.
\cite{PW88}.
Moreover it has been suggested that the discrepancy may
be explained through the effect of the IR cutoff
\cite{KN92}. 
However the analyses \cite{KEIT95,KM95} based on the
nonlocal gauge have confirmed most of the original results
\cite{ABKW86}. Among other things, they proved that there
exists a finite critical number of flavor
$N_f^c \cong 128/(3\pi^2) \cong 4.32$ below which
($N_f<N_f^c$) the chiral symmetry is spontaneously broken
and the fermion mass is dynamically generated, and that
 the scaling law is given by the essential singularity type
(at the critical point $N_f^c$) for the dynamical fermion
mass and the chiral order parameter
$\langle \bar \Psi \Psi \rangle$.
The resulting value of $N_f^c$ agrees with that of the
improved calculation incorporating $1/N_f^2$ corrections by
Nash
\cite{Nash89}. Note that the above result in the
neighbourhood of the critical point does not depend on the
explicit choice of $G$, as long as the vertex is consistent
with the Ward-Takahashi (WT) identity, see \cite{Kondo97}.

\appendix
\section{Kernel}
  By changing the integration variable from $\theta$ to
$r:=\sqrt{k^2}=\sqrt{p^2 +q^2 - 2pq \cos \theta}$,
the kernel
(\ref{3dK})  can be rewritten as
\begin{eqnarray}
 4\pi^2 L_3(p,q) = {1 \over 2pq} \int_{|p-q|}^{p+q} dr
 (p^2-q^2-r^2)r f(r) .
\end{eqnarray}
Integrating the RHS term by term using the expansion 
(\ref{expandf}) of $f(k)$, we obtain
\begin{eqnarray}
 4\pi^2 L_3(p,q) &=& {1 \over 2pq} 
 \Biggr[ - 2 \sum_{n=0}^{\infty} {(-1)^n \over n+3}
 {(p+q)^{n+1} - |p-q|^{n+1} \over \alpha^{n+1}}
 \nonumber\\
 &&+ 2(p^2-q^2) \sum_{n=0(n\not=1)}^{\infty}
 {(-1)^n (n+1) \over  (n-1)(n+3)} 
 {(p+q)^{n-1} - |p-q|^{n-1} \over \alpha^{n+1}}
 \nonumber\\
 &&- {p^2-q^2 \over \alpha^2} \ln {p+q \over |p-q|}
 \Biggr] .
 \label{kernelexpand}
\end{eqnarray}
Here note that the last logarithmic term corresponds to
the $n=1$ case which is ommited in the summation in the
second term on the RHS of the above equation.
Up to $n=0$ or $n=1$, this reproduces (\ref{3dK1}) or 
(\ref{3dK2}), respectively.
\par
On the other hand, using (\ref{fk}) for $f(k)$, we arrive
at the complete kernel:
\be
4\pi^2 L_3(p,q) &=&
 {\alpha(p^2-q^2) \over 2pq}
 \left[{1 \over |p-q|^3}-{1 \over (p+q)^3} \right]
\nonumber\\&&
-  {p^2-q^2 \over 4pq}
 \left[{1 \over |p-q|^2}-{1 \over (p+q)^2} \right]
\nonumber\\&& 
-  {2\alpha^2-p^2+q^2 \over 2\alpha pq}
 \left[{1 \over |p-q|}-{1 \over (p+q)} \right]
 \nonumber\\
&&+ {p^2-q^2 \over 2\alpha^2 pq} \ln \left[
 {\alpha+p+q \over \alpha+|p-q|}  {|p-q| \over p+q} \right]
 \nonumber\\
&&-  {\alpha^2 \over 2pq} \left[
   {p+3q \over (p+q)^3} \ln {\alpha+p+q \over \alpha}
 - {p-3q \over (p-q)^3} \ln {\alpha+|p-q| \over \alpha}
\right] .
 \label{kernelL}
\end{eqnarray}
Indeed, this agrees with (\ref{kernelexpand}) after
expanding (\ref{kernelL}) in powers of $1/\alpha$.
\par
This should be compared with the kernel in the SD equation
(\ref{SDBnlg}) for $B$ in the nonlocal gauge.
\begin{eqnarray}
  g(k) := D_T(k^2)[2+\tilde \xi(k^2)]
= {4 \over k^2 + \alpha k} - {2 \over k^2} \left[
  1 - {2\alpha \over k} + {2\alpha^2 \over k^2} 
  \ln \left( 1+{k \over \alpha} \right) \right] .
  \label{gk}
\end{eqnarray}
This leads after angular integration to the kernel
\cite{KM95}:
\begin{eqnarray}
 8\pi^2 K_3(p,q) &=&  
 {\alpha \over pq}{p+q-|p-q| \over |p-q|(p+q)}
 + {1 \over pq} \ln {\alpha+p+q \over \alpha+|p-q|}
 \nonumber\\
&&+  {\alpha^2 \over pq} \left[
   {1 \over (p+q)^2} \ln {\alpha+p+q \over \alpha}
 - {1 \over |p-q|^2} \ln {\alpha+|p-q| \over \alpha}
\right] .
 \label{kernelK}
\end{eqnarray}
If we use the expansion
\begin{eqnarray}
 g(k)= \sum_{n=0}^{\infty} 4 (-1)^n {n+2 \over n+3} 
  {k^{n-1} \over \alpha^{n+1}} 
  =  {8 \over 3\alpha k} - {3 \over \alpha^2}
  + O(k) ,
  \label{expandg}
\end{eqnarray} 
the kernel (\ref{kernelK}) reads \cite{KEIT95}
\begin{eqnarray}
 8\pi^2 K_3(p,q) = {1 \over 2pq} \int_{|p-q|}^{p+q} dr
r g(r) 
&=&  {4 \over pq} 
  \sum_{n=0}^{\infty} {(-1)^n (n+2) \over (n+1)(n+3)}
 {(p+q)^{n+1} - |p-q|^{n+1} \over \alpha^{n+1}}
 \nonumber\\ 
&=& {8 \over 3\alpha} \left[ {p+q-|p-q| \over 2pq}
- {9 \over 8} {1 \over \alpha^2} \right] + O(1/\alpha^3) .
\end{eqnarray}
It has been analytically shown \cite{KEIT95} that the
inclusion of the next-to-leading term in the kernel does
not change the critical behavior near the critical number
of flavors.  Subsequently, this has been confirmed by the
numerical study using the complete expression
(\ref{kernelK}) in Ref.~\cite{KM95}.

%\newpage

\vskip 1cm
\centerline{\large Figure Caption}

\vskip 0.5cm

Fig.1:  Wave function renormalization $Z(p)$ as a function
of $p/\alpha$ for the leading (\ref{3dK1}),
the next-to-leading (\ref{3dK2}) and the full kernels
(\ref{kernelL}) when $N_f=5$.

\end{document}